\documentstyle[12pt,amsbsy]{article}
\newcommand{\be}[1]{\begin{equation}\label{#1}}
\newcommand{\ee}{\end{equation}}
\newcommand{\beq}[1]{\begin{eqnarray}\label{#1}}
\newcommand{\eeq}{\end{eqnarray}}
\newcommand{\ba}{\begin{array}}
\newcommand{\ea}{\end{array}}
\newcommand{\of}[1]{\left(#1\right)}
\newcommand{\off}[1]{\left[#1\right]}
\newcommand{\offf}[1]{\left\{#1\right\}}
\newcommand{\bs}[1]{\boldsymbol{#1}}
\newcommand{\ms}[1]{\mbox{\scriptsize{#1}}}

\newcommand{\Par}{\par\vspace{0.2cm}\par}

\newcommand{\A}{\bar{a}}
\newcommand{\B}{\bar{b}}
\newcommand{\C}{\bar{c}}

\newcommand{\N}{\bar{n}}
\newcommand{\LL}{\bar{L}}
\newcommand{\Alpha}{\bar{\alpha}}
\newcommand{\Beta}{\bar{\beta}}
\newcommand{\GGamma}{\bar{\gamma}}

\newcommand{\bp}{\Par\noindent{\em\underline{Proof\/}}:$\;\;$}
\newcommand{\ep}{\hfill $\Box$\Par\noindent}
\newcommand{\bP}{\Par\noindent{\em\underline{Proposition\/}}:$\;\;$}
\newcommand{\eP}{\Par\noindent}
\newcommand{\ac}{\mathaccent"7017}
\tiny\normalsize
\begin{document}
\sloppy
\title{\Huge {\sf On the Composition of Gauge Structures}}
\author{{\normalsize\sc OFER MEGGED}\thanks{e-mail address:
megged@ccsg.tau.ac.il}\\
{\normalsize\rm School of Physics and Astronomy}\\
{\normalsize\rm Tel-Aviv University, Tel-Aviv 69978, Israel.}}
\date{January 9, 1996}
\maketitle
\begin{abstract}\noindent
A formulation for a non-trivial composition of two classical gauge 
structures is given: Two parent gauge structures of a common base 
space are synthesized so as to obtain a daughter structure which 
is fundamental by itself. The model is based on a pair of related 
connections that take their values in the product space of the 
corresponding Lie algebras. The curvature, the covariant exterior 
derivatives and the associated structural identities, all get 
contributions from both gauge groups. The various induced
structures are classified into those whose composition is 
given just by trivial means, and those which possess an irreducible 
nature. The pure irreducible piece, in particular, generates a complete
super-space of ghosts with an attendant set of super-BRST variation
laws, both of which are purely of a geometrical origin.
\end{abstract}
\newpage\voffset -1.2cm
\setlength{\textheight}{19.4cm}
\section{\sf{Introduction}}\label{s1}
It has generally been accepted that composite classical gauge structures 
of a non-supersymmetric nature are constructed via trivial splicing. 
Indeed, the formal description of principal bundles could thus easily 
be extended to include bundles whose fibers are product spaces, with a 
different gauge-group acting independently on each factor fiberspace
in the product. Any other composition process breaks this 
formal scheme within which different gauge structures, introduced on the 
same underlying manifold, are mutually transparent to one another and 
might somehow be correlated only by artificial means.\par   
This note suggests a new type of gauge theory which results from a non-trivial 
composition of two genuine gauge structures. In this theory the
geometry is not split even though the bundle is.
It is first developed for 
splices of fiber bundles which comply with some severe algebraic 
requirements realizing very clear-cut principles.
Extending the theory to incorporate a wider class 
of gauge structures is shown to be straightforward, provided some soft 
structural requirements at the level of the algebra are fulfilled. 
Within the extended framework, the single-fiber structures and 
the process of trivial splicing appear only as sub-sectors 
in a much larger and comprehensive construction.\par
In what follows, things are formulated in a geometrical setting by using 
a coordinate-free language. After a short prelude which serves mainly to 
fix the notation, we define the notion of a foliar complex, exhibiting
its rich geometrical content and ``peculiar'' algebraic structure. In
particular, the covariant bundle operators and the associated
structural identities are strictly recovered. We then present an 
algebraic interpretation, with which more insight is gained and geometric 
links are drawn. The invariance of the curvature with respect to local 
translations in the spaces of connections is later examined and it is 
found that it directly implies a whole super-BRST sector. Finally, upon 
removing some of the initial constraining requirements, the theory is 
shown to accommodate many types of gauge structures.
\section{\sf{Foliar Complexes on Spliced fiber bundles}}\label{s2}
Consider a smooth manifold $M$ of $n$ dimensions, defined over a field 
$K$. Let $T_{x}M$ be the space tangent to $M$ at a point $x\in M$, and
let $T_{x}^{\star}M$ denote the space of all linear functionals on $T_{x}M$.
One usually constructs the {\bf classical Grassmann bundle} $\wedge TM$,
the bundle whose sections are anti-symmetric tensor fields on $M$, and the
counterpart co-bundle $\wedge T^{\star}M$, the one whose sections are
$K$-valued differential forms, by gluing together the Grassmann algebras
associated with each and every (co-) tangent space at each and every
point of $M$,
\be{a}
\wedge TM \,:=\, \bigcup_{x\in M}\of{\bigoplus_{p=1}^{n}\bigwedge\!\!{}^{p}
T_{x}M},\;\;\;\; \wedge\, T^{\star}M \,:=\, \bigcup_{x\in
M}\of{\bigoplus_{p=0}^{n} \bigwedge\!\!{}^{p} T_{x}^{\star}M},
\ee
where the union and the direct sum are commutable. In particular, 
the first summand of $\wedge TM$ above is the tangent bundle, the zero-th 
summand of $\wedge T^{\star}M$ is the space of all functions on $M$, and
$T^{\star}M$ is the co-tangent bundle of one-forms. For any $\alpha,\beta$ 
differential forms $\in\wedge T^{\star}M$, one defines their graded brackets
(``commutator'') by $\off{\alpha,\beta}:=\alpha\wedge\beta
-\of{-1}^{pq}\beta\wedge\alpha$, where $p$ is the form-degree of $\alpha$ 
and $q$ is that of $\beta$. In addition, the underlying manifold 
is naturally equipped with an exterior derivative $d$ which maps $p$-graded 
objects into $\of{p+1}$-graded ones. Being a co-boundary operator over 
$\wedge T^{\star}M$, it satisfies $d\circ d=0$ on forms, and obeys the 
graded Leibnitz rule with respect to wedge multiplication: 
$d\of{\alpha\wedge\beta}=d\alpha\wedge\beta+\of{-1}^{p}\alpha\wedge 
d\beta$.\Par 
Once a structure group and a representation space are specified, a gauge 
structure is established. The {\bf classical gauge} is by definition a 
$\mbox{GL}\of{n,K}$ structure in its fundamental representation, that is, 
the gauge group is taken to be the group of general linear transformations 
acting on the fibers of the classical Grassmann bundle. The classical gauge 
directly decodes many of the geometric properties of $M$ and, with the 
additional requirement of being invariant under base space diffeomorphisms, 
provides a framework for $n$-dimensional theory of general
relativity (without spinors).
In its very-nature the construction is self-symmetric: 
base-space components and fiber components are completely interchangeable, 
and quite often mutually contracted.\Par
Consider next two finite dimensional representation spaces      
$V$ and $U$, of two respective distinct Lie groups $G_{V}$ and $G_{U}$,
with respective dimensions $n_{V}$ and $n_{U}$, each of which meet the
following stringent requirement: each set of generators of the two forming
groups, when carrying an appropriate representation in the corresponding
representation space, must \underline{close} under anti-commutation 
relations as well \cite{2}. The existence
of these closure requirements, as well as their formal form, crucially 
depends on the particular representation that is chosen. For example, it 
may be satisfied in a fundamental representation, but may
not exist, or exist in a different form, for higher representations. In
particular $V$ and $U$ are taken to be the fibers of two associated bundles
having a common base space, a smooth manifold $M$ of $n$ dimensions. Because
the two associated bundles $VM$ and $UM$ are smooth by themselves, the
so-called {\bf local leaves} $S_{x}:=V_{x}\times U_{x}$ can be smoothly
glued to form a spliced fiber bundle $SM:=\bigcup_{x\in M}S_{x}$ composed 
of two structures. In a similar manner, one may take copies of the product
space $V_{x}\times U_{x}$, and form two-structure bundles of higher
dimensional leaves.\Par
Once the splice is globally formed, $x$-dependent frames are assigned to
each and every local leaf. The smoothly-glued local frames draw sections 
(frame fields)
with which every geometric object $\in SM$ can be described. In what
follows we shall draw the attention only to those geometric differential
forms on which $G_{V}$ and $G_{U}$ act linearly, and in the same manner.
Termed as {\bf leaf-valued forms}, they transform like leaf vectors 
or leaf rank-$r$ tensors with respect to the combined action of 
$G_{V}\times G_{U}$, but we shall not restrict ourselves only to 
combined actions. Another type of differential forms with which we are 
much interested are forms of the structural kind, taking their values 
in the \underline{product-space} of the corresponding two Lie algebras, 
$\mbox{Lie}\,G_{V}\otimes\mbox{Lie}\,G_{U}$. The whole of leaf-valued
differential forms and the whole of $\of{\mbox{Lie}\otimes\mbox{Lie}}$-valued 
ones constitute a totality of forms $\Gamma_{FC}$, each of its members 
represented irreducibly and non-trivially on both fibers. The corresponding
{\bf foliar complex} $FC$ of irreducible bundle objects will just be
associated with a particular geometric infrastructure of gauge in
which projections on fibers, and the global process of identity
reduction, are no longer natural attributes of the
splice. In fact, the foliar formation accommodates two group 
structures in a non-contractable fashion.
\Par\noindent{\em Comment\/}:
The obligatory demand for closure of the algebras under anti-commutation 
can be much softened, for example, by considering cases where 
anti-commutators of group generators in a representation 
include {\em as well\/} terms proportional to the identity. 
This type of extension necessarily generalizes
the concept of a foliar complex. For pedagogical reasons, however, we 
postpone any of this to section \ref{s6}.
\section{\sf{The Geometry Associated with a Foliar Complex}}\label{s3}
Let us begin by proposing the following:
\bP\label{P1}
The two-form structure
\be{1}\ba{l}
R_{FC}\,:=\:d\of{\varphi+\omega}+\of{\varphi+\omega}\wedge\of{\varphi+\omega}
\\=\;d\varphi+\varphi\wedge\varphi+\varphi\wedge\omega+\omega\wedge\varphi
+\omega\wedge\omega+d\omega
\ea\ee
is a linear curvature $\in FC$ provided that the pair of connection
one-forms $\of{\omega,\varphi}$ take their values in \underline{$\mbox{Lie}
\,G_{V} \otimes \mbox{Lie}\,G_{U}$} and obey the set of transformation laws
\be{2}  \ba{lll}
\forall\, v \in G_{V}: & \omega\mapsto v \omega v^{-1}, &
\varphi\mapsto v \of{\varphi + d} v^{-1},  \\
\forall\, u \in G_{U}: & \varphi\mapsto u \varphi u^{-1}, &
\omega\mapsto u \of{\omega + d} u^{-1}; \ea
\ee
that is, the interleaved fibers interchange the geometric roles played by
$\omega$ and $\varphi$: from the point of view of $G_{V}$, $\varphi$ plays
the role of a connection and $\omega$ transforms as a tensor. However,
from the point of view of $G_{U}$, $\omega$ plays the role of a
connection and $\varphi$ transforms as a tensor.
\eP
{\em Comment\/}: True, $\mbox{Lie}\,G_{V}\otimes\mbox{Lie}\,G_{U}
\neq \mbox{Lie}\of{G_{V}\times G_{U}} = \mbox{Lie}\,G_{V}\otimes I_{U}+
I_{V}\otimes\mbox{Lie}\,G_{U}$, where $I$ stands for an identity. 
Please notice, however, that $\mbox{Lie}\,G_{V}\otimes\mbox{Lie}\,G_{U}$ 
and $\mbox{Lie}\of{G_{V}\times G_{U}}$ are both embedded in
$\of{\mbox{Lie}\,G_{V}+I_{V}}\otimes\of{\mbox{Lie}\,G_{U}+I_{U}}$.
\bp
The proof follows two steps: one first shows that the coefficients of the 
wedge-products in $R_{FC}$ lie in $\mbox{Lie}\,G_{V}\otimes\mbox{Lie}\,G_{U}$, 
and second, that $R_{FC}$ transforms linearly, and in an independent manner,
with respect to both gauge groups. Throughout this work we use bare letters 
to denote anything which is attached to the $V$-fiber and barred letters to 
denote anything which is attached to the $U$-one. Let then the 
$n_{V}\times n_{U}$ $\of{=\mbox{dim}\,G_{V}\times\mbox{dim}\,G_{U}}$ 
tensor products $L_{a}\otimes\LL_{\A}$ span a basis for 
$\mbox{Lie}\,G_{V}\otimes\mbox{Lie}\,G_{U}$ in the leaf representation
space $S=V\times U$. In this basis the two foliar connections read $\omega 
=\omega_{a\A}\,L^{a}\otimes\LL^{\A}$ and $\varphi=\varphi_{a\A}\,L^{a}
\otimes\LL^{\A}$. The realization of the commutation relations among 
the generators must be followed (formally at least) by the indispensable 
requirement for closure anticommutability:
\be{3}\ba{lcl}
\off{L^{a},L^{b}}\,=\,f_{V}^{abc}L_{c} &  &
\offf{L^{a},L^{b}}\,=\,g_{V}^{abc}L_{c} \\
\off{\LL^{\A},\LL^{\B}}\,=\,f_{U}^{\A\B\C}\LL_{\C} &  &
\offf{\LL^{\A},\LL^{\B}}\,=\,g_{U}^{\A\B\C}\LL_{\C}\,.
\ea\ee
Note: The algebra's {\em structure constants\/} $f$, and 
the {\em Clebsch-Gordan coefficients\/} $g$ of the 
$V\times V\mapsto V$ and $U\times U\mapsto U$ homomorphisms, 
are not automatically derived form trace formulae since the 
associated Cartan metric tensors are usually no longer invertible. 
Now, the term $\omega\wedge\omega$, say, reads
(square-brackets over indices stand for anti-symmetrization; 
representation space indices are suppressed):
\beq{4}
\omega\wedge\omega &\!\!=\!\!& \omega_{\,[\mu| a\A}
L^{a}\otimes\LL^{\A}\omega_{\nu]\,b\B}
L^{b}\otimes\LL^{\B}\otimes\bs{e}^{\mu}\wedge\bs{e}^{\nu} \nonumber\\
&\!\!=\!\!& \omega_{\mu\, a\A\,}\omega_{\nu\, b\B}\off{L^{a}\otimes\LL^{\A},
L^{b}\otimes\LL^{\B}}\otimes\bs{e}^{\mu}\wedge\bs{e}^{\nu},
\eeq
where the set of $n$ $\of{=\mbox{dim}\,M}$ co-frame fields 
$\offf{\bs{e}^{\mu}}$ span the basis for $T^{\star}M$. However, 
$\off{L^{a}\otimes\LL^{\A},L^{b}\otimes\LL^{\B}}=
L^{a}L^{b}\otimes\off{\LL^{\A},\LL^{\B}}+\LL^{\A}\LL^{\B}\otimes
\off{L^{a},L^{b}}-\off{L^{a},L^{b}}\otimes\off{\LL^{\A},\LL^{\B}}$.
The third term of the resulting expansion clearly lies in $\mbox{Lie}\,G_{V}
\otimes\mbox{Lie}\,G_{U}$. As for the other two terms in the expansion,
one should first reconstruct the wedge multiplication and then, only with 
respect to one of the anti-symmetric counterparts, simultaneously re-shuffle 
summed bare and barred algebra-space indices. As a result, and due to the 
anti-symmetry of the algebra structure constants in their first two indices, 
the two free pairs of generators respectively convert to anti-commutators, 
$L^{a}L^{b}\rightarrow\offf{L^{a},L^{b}}$,
$\LL^{\A}\LL^{\B}\rightarrow\offf{\LL^{\A},\LL^{\B}}$, and the closure
conditions (\ref{3}) can then be fully implemented. The product of
$\omega \wedge\omega$ finally reads $\frac{1}{2}C_{VU\;c\C}^{a\A b\B}\,
\omega_{a\A}\wedge\omega_{b\B}L^{c}\otimes\LL^{\C}$, where we introduce the
{\bf foliar-constants} $C_{VU}^{a\A b\B c\C}\equiv\frac{1}{2}\of
{f_{V}^{abc} g_{U}^{\A\B\C} + f_{U}^{\A\B\C} g_{V} ^{abc}} - f_{V}^{abc}f_{U}
^{\A\B\C}$, characteristic of foliar complexes. The considerations above
certainly apply to $\varphi\wedge\varphi$, and moreover, to 
$\off{\varphi,\omega}=\varphi\wedge\omega+\omega\wedge\varphi$, 
always along with the same $C_{VU}$'s. Therefore,
\be{100}
R_{FC}\;=\sum_{\varpi,\varpi'=\varphi,\omega}
\of{d\varpi_{c\C}+\frac{1}{2}C_{VU\;\;c\C}^{a\A b\B}
\varpi_{a\A}\wedge\varpi'_{b\B}}
L^{c}\otimes\LL^{\C}.
\ee
This concludes our analysis 
of the structure of the curvature, which is thus seen to lie in
$\mbox{Lie}\,G_{V}\otimes\mbox{Lie}\,G_{U}$.\par
One's second concern corresponds to the transformation properties of 
$R_{FC}$ $\of{\equiv R}$. Consider first the action applied to the 
curvature by an element $v$ of $G_{V}$. Making use of 
$vdv^{-1}=-\of{dv}v^{-1}$ and taking care of the correct signs in 
successive operations (for example, $d\circ\varphi\wedge\of{\cdot}= 
d\varphi\wedge\of{\cdot}-\varphi\wedge d\of{\cdot}$) 
one immediately infers that $R$ can be decomposed into three terms, 
$d\omega+\varphi\wedge\omega+\omega\wedge\varphi$ and $\omega\wedge
\omega$ and $d\varphi+\varphi\wedge\varphi$, each of which transforms
linearly in an independent manner. Acting, however, with an element $u$
of $G_{U}$, one finds a different decomposition of linearly transformed
terms, obtained from the previous one by the interchange
$\omega\leftrightarrow\varphi$. Therefore, $\forall\,v\in G_{V}$,
$R\mapsto v R v^{-1}$, and $\forall\,u\in G_{U}$, $R\mapsto u R u^{-1}$, 
and because $G_{V}$ and $G_{U}$ act in different spaces, we also have 
$\forall\,\of{v\times u}\in G_{V}\times G_{U}$, $R\mapsto \of{v\times u}
R\of{v\times u}^{-1}$.\ep\indent
The pair of connection one-forms and the associated curvature two-form
are structure-type forms $\in\Gamma_{FC}$. They take their values in a 
space spanned by constant matrices. In what follows we would like to 
develop the concept of covariant differentiation of leaf-valued 
forms, sections of the spliced fiber bundle:
Let $\psi$ denote a generic leaf $p$-form, $G_{V}\times G_{U}$ 
vector-valued, or $G_{V}\times G_{U}$ tensor-valued of an arbitrary rank.
Then, if $\psi$ is vector-valued, so is the quantity
$D\psi := d\psi + \of{\omega + \varphi}\wedge\psi$ and if $\psi$ is
tensor-valued, $D^{\star}\psi := d\psi + \of{\omega + \varphi}\wedge\psi
+ \of{-1}^{\of{p+1}}\psi\wedge\of{\omega + \varphi}$ is tensor-valued as
well. The two operators $D$ and $D^{\star}$ (whose powers are defined
through composition) are structure-preserving differentiations of the 
foliar complex. They are respectively called the {\bf covariant exterior 
derivatives} of vector-valued and tensor-valued foliar forms. Notice, however, 
that the application of the graded Leibnitz rule in the context of these 
exterior derivatives, only makes sense for homogeneous multiples of 
leaf-valued forms. Namely, all forms in the product are one by one 
leaf vector-valued or else, one by one leaf tensor-valued. Otherwise, 
the resulting derivation will transform non-linearly. It is exactly
for this reason that homogeneous compositions are so fundamental to 
foliar complexes; the non-homogeneous ones do not support covariance.\Par
Having the covariant exterior differentiations in hand, one may redefine the 
two-form curvature via $\off{D,D}\psi = 2\of{d+\of{\omega+\varphi}\wedge}
\circ\of{d+\of{\omega+\varphi}\wedge}\psi:=2R\wedge\psi$, or alternatively
via $\off{D^{\star},D^{\star}}\psi = 2D^{\star}\circ D^{\star}\psi=
2\of{R\wedge\psi-\psi\wedge R}:=2\off{R,\psi}$. Now, if one is able to
find such a special form $\psi=\Psi$, valued in 
$\mbox{Lie}\,G_{V}\otimes\mbox{Lie}\,G_{U}$, then the second definition 
above is just the Bianchi-like identity for the foliar's {\em torsion\/} 
$T:=D^{\star}\Psi$. Furthermore, by a direct calculation, or by applying
the Jacobi identity, $0=\off{D^{\star},
\off{D^{\star},D^{\star}}}\psi=2D^{\star}\circ\off{R,\psi}-2\off{R,
D^{\star}\psi}$, one arrives at $D^{\star}R = 0$ which is
the Bianchi-like identity for the curvature. These two structural
identities in turn imply the Ricci identity $D^{\star}\circ D^{\star}
T=\off{R,T}$. Moreover, let us put $\phi:=D^{\star}\psi$. Then, for
complexes over infinite dimensional manifolds (with finite dimensional
leaves), the asymptotic formula $\of{\exp
D^{\star}\circ D^{\star}} \circ\phi=\of{\exp R}\phi\of{\exp -R}$ applies.
This can be vividly seen by noting that the action of the $p$-th power
of $D^{\star}\circ D^{\star}$ on $\phi$ produces a $p$-nested even
commutator of the type $\off{R,\off{R,\off{\cdots,\off{R,\phi}\cdots}}}$
which is, up to a factorial prefactor, the $p$-th term in the well-known
formula for induced representations. Yet, the identification $\phi=T$
is somewhat problematic since in this case the underlying leaf carries
infinite dimensions. 
\section{\sf{Algebraic Formulation of a Foliar Structure}}\label{s4}
The two connections associated with the foliar complex can also
be defined algebraically by means of infinitesimal changes in foliar
frames. Let $X$ denote an arbitrary vector-field on $M$ and let the
$N_{V}\times N_{U}$
$\of{=\mbox{dim}\,V\times\mbox{dim}\,U}$ frame-fields $\bs{e}_{a}^{A}\otimes
\bar{\bs{e}}_{\A}^{\bar{A}}$ span a basis for a leaf $S$ at each point of
$M$. Here and through the rest of this section small letters from the 
beginning of the alpha-bet label basis vectors, the corresponding capital
letters label their components. 
The differential of a fiber basis vector
is clearly linear in that basis and given in
terms of the connection one-form by the defining set of equations
$d\bs{e}_{a}^{A}=\widetilde\varphi_{a}^{b}\of{X}\bs{e}_{b}^{A}:=
-\varphi_{m}\of{X}{L^{m}}^{A}_{B}\bs{e}_{a}^{B}$, where the
index $m$ (and later on, $\N$ as well) runs over the algebra,
and the coefficients are evaluated on $X$. The inclusion
of a second fiber within the context of a foliar complex can only be
consistently done by considering the coefficients $\widetilde\varphi$
as tensor-valued 
in the counterpart representation space, $\of{d\bs{e}_{a}^{A}}\otimes
\bar{\bs{e}}_{\A}^{\bar{A}}={{\widetilde\varphi}_{a\bar{B}}}^{b\bar{A}}
\of{X}\bs{e}_{b}^{A}\otimes\bar{\bs{e}}_{\A}^{\bar{B}}:=
-\varphi_{m\N}\of{X}{L^{m}}^{A}_{B}\bs{e}_{a}^{B}\otimes
\LL^{\,\N\bar{A}}_{\;\:\bar{B}}\bar{\bs{e}}_{\A}^{\bar{B}}$,
where the $n_{U}$ $\LL$'s are the generators of the counterpart
group. Taking off the indices we write $\of{d\bs{e}}\otimes\bar{\bs{e}}
=-\varphi\of{\bs{e}\otimes\bar{\bs{e}}}$ where it is understood that
$\varphi$ is a one-form $\in T^{\star}M$ whose action on a leaf basis
$\of{\bs{e}\otimes\bar{\bs{e}}}$
is carried-out via an appropriate representation of the generators.
Now, for the computation of the differential of the entire basis, 
two distinct connections must be (uniquely) introduced:
\be{18}\ba{lcl}
d \of{\bs{e}_{a}^{A}\otimes\bar{\bs{e}}_{\A}^{\bar{A}}}
&\!\!\!=\!\!\!& \of{d \bs{e}_{a}^{A}}\otimes\bar{\bs{e}}_{\A}^{\bar{A}} +
\bs{e}_{a}^{A}\otimes \of{d \bar{\bs{e}}_{\A}^{\bar{A}}}
=\; \widetilde\varphi_{a\bar{B}}^{b\bar{A}}\of{X}
\bs{e}_{b}^{A}\otimes\bar{\bs{e}}_{\A}^{\bar{B}} +
\widetilde\omega_{\A B}^{\B A}\of{X}
\bs{e}_{a}^{B}\otimes\bar{\bs{e}}_{\B}^{\bar{A}} \\
&\!\!\!:=\!\!\!& -\varphi_{m\N}\of{X}{L^{m}}^{A}_{B}\bs{e}_{a}^{B}\otimes
\LL^{\,\N\bar{A}}_{\;\:\bar{B}}\bar{\bs{e}}_{\A}^{\bar{B}}-
\omega_{m\N}\of{X}{L^{m}}^{A}_{B}\bs{e}_{a}^{B}\otimes
\LL^{\,\N\bar{A}}_{\;\:\bar{B}}\bar{\bs{e}}_{\A}^{\bar{B}},
\ea\ee
that is,\, $d\of{\bs{e}\otimes\bar{\bs{e}}}=-\varphi\of{\bs{e}\otimes
\bar{\bs{e}}}-\omega\of{\bs{e}\otimes\bar{\bs{e}}}$.
Despite the structural resemblance in the two terms above,
the coefficients of $\varphi$ and $\omega$ are indeed different entities;
whereas $\varphi$ is uniquely determined by the set of equations which
relates it to $\widetilde\varphi$, which is induced by the gauge-group
$G_{V}$, $\omega$ is determined by a different set of equations, namely,
the one which relates it to $\widetilde\omega$, induced by the gauge-group
$G_{U}$. Of course, the transformation laws of (\ref{2}) are solutions of 
definition (\ref{18}); but furthermore, consistency requires the 
compatibility of (\ref{18}) with a {\em combined\/} $G_{V}$-$G_{U}$ 
$\of{=\mbox{$G_{U}$-$G_{V}$}}$ gauge transformation on the entire leaf.
Let us see that this is indeed the case:
An action of $\of{v\times u}\in G_{V}\times G_{U}$ to the left of 
the leaf basis reads: $\of{v\times u}\of{\bs{e}\otimes\bar{\bs{e}}}:=
\of{v\bs{e}\otimes u\bar{\bs{e}}}$. Now, for the differential of a 
transformed basis we get,
\be{p2a} 
d\of{v\bs{e}\otimes u\bar{\bs{e}}} =
\of{dv}\bs{e}\otimes u\bar{\bs{e}}-v\circ\varphi'\of{\bs{e}\otimes
u\bar{\bs{e}}}+v\bs{e}\otimes\of{du}\bar{\bs{e}}
-\of{v\times u}\circ\omega\of{\bs{e}\otimes\bar{\bs{e}}},
\ee
where $\varphi'$ (which is $\varphi$ transformed by $G_{U}$) is still 
left to be determined. On the other hand, the transformed 
left-hand side of (\ref{18}) reads:
$G_{V}\times G_{U}:\,-\varphi\of{\bs{e}\otimes\bar{\bs{e}}}-\omega
\of{\bs{e}\otimes\bar{\bs{e}}}\,\mapsto\,
-\varphi''\of{v\bs{e}\otimes u\bar{\bs{e}}}-\omega''
\of{v\bs{e}\otimes u\bar{\bs{e}}}$.
Following the laws of (\ref{2}) we
set $-\varphi''=-u\of{v\varphi v^{-1} +vdv^{-1}}u^{-1}$ and
$-\omega''=-v\of{u\omega u^{-1} +udu^{-1}}v^{-1}$ from which
\be{p2c}\ba{l}
-\varphi''\of{v\bs{e}\otimes u\bar{\bs{e}}} =
-\of{u\times v}\circ\varphi\of{\bs{e}\otimes\bar{\bs{e}}} + 
\of{dv}\bs{e}\otimes u\bar{\bs{e}}\\
-\omega''\of{v\bs{e}\otimes u\bar{\bs{e}}} =
-\of{v\times u}\circ\omega\of{\bs{e}\otimes\bar{\bs{e}}}
+v\bs{e}\otimes\of{du}\bar{\bs{e}}.
\ea\ee
Identifying now the two left-hand sides of (\ref{p2c}) with the left-hand 
side of (\ref{p2a}) implies $\varphi'\of{\bs{e}\otimes u\bar{\bs{e}}}
=u\circ\varphi\of{\bs{e}\otimes\bar{\bs{e}}}$, in precise agreement with 
$G_{U}:\varphi\mapsto u\varphi u^{-1}$.\Par
{\em Example\/}:
In a fundamental representation, the indices which label the basis vectors
are of the same type as those which label their components.
In particular, for the classical groups (and their related sub-structures)
the algebra as well employs the same type of indices. Consider 
a Whitney product of two tangent bundles, both associated with 
the same manifold $M$. The arena of leaf-valued forms lies in 
\[
\bigcup_{x\in M}
{\of{\bigoplus_{p=0}^{n}\bigwedge\!\!{}^{p}T_{x}^{\star}M}}
{\of{\bigotimes_{\alpha=1}^{2}T^{\of{\alpha}}_{x}M}}.
\]
The foliar structure is induced by two independent 
$\mbox{GL}\of{n,R}$ groups whose generating elements 
are realized on $T_{x}M$ via the defining representation  
$(L^{a}_{b})^{B}_{A}=\delta^{a}_{A}\delta^{B}_{b}$,
from which commutation $\of{-}$ and anti-commutation $\of{+}$ 
relations are easily computed: $\off{L^{a}_{b},L^{c}_{d}}_{\mp}=
\of{\delta^{c}_{\,b}\delta^{a}_{\,e}\delta^{f}_{\,d}\mp
\delta^{a}_{\,d}\delta^{c}_{\,e}\delta^{f}_{\,b}}L^{e}_{f}$.
The realizations above imply that $\varphi$ and $\omega$, both acquire 
a particularly simple form: $\varphi^{a\A}_{b\B}({L^{b}_{a}
\otimes L^{\B}_{\A}})^{A\bar{A}}_{B\bar{B}}=\varphi^{A\bar{A}}_{B\bar{B}}
=-\tilde{\varphi}^{A\bar{A}}_{B\bar{B}}$ and  
$\omega^{a\A}_{b\B}({L^{b}_{a}\otimes L^{\B}_{\A}})^{A\bar{A}}_{B\bar{B}}
=\omega^{A\bar{A}}_{B\bar{B}}=-\tilde{\omega}^{A\bar{A}}_{B\bar{B}}\,$.
The classical gauge thus identifies the notion of parallelism,
which is based on the concept of a tilde-connection, with that of 
horizontality, based on Yang-Mills connections with values in 
Lie algebras. Consequently, 
$R^{A\bar{A}}_{B\bar{B}}= d\varphi^{A\bar{A}}_{B\bar{B}}+
\varphi^{A\bar{A}}_{C\bar{C}}\wedge\varphi^{C\bar{C}}_{B\bar{B}}+
\varphi^{A\bar{A}}_{C\bar{C}}\wedge\omega^{C\bar{C}}_{B\bar{B}}+
\omega^{A\bar{A}}_{C\bar{C}}\wedge\varphi^{C\bar{C}}_{B\bar{B}}+
\omega^{A\bar{A}}_{C\bar{C}}\wedge\omega^{C\bar{C}}_{B\bar{B}}+
d\omega^{A\bar{A}}_{B\bar{B}}\,$ which can be viewed as a 
generalization of the expression 
for the components of a single-structure curvature, given by 
$R^{A}_{B}=d\varphi^{A}_{B}+\varphi^{A}_{C}\wedge\varphi^{C}_{B}$.
Indeed, the {\bf folium curvature\/} of a smooth manifold $M$, 
is given by a $\of{2,2}$-tensor two-form on $M$.\Par
On the basis of the above viewpoints, one is naturally led to infer that 
the curvature $R_{FC}$ of a foliar complex correlates between the two 
``half-linear'' primordial curvatures $d\varphi+\varphi\wedge\varphi$ 
and $d\omega+\omega\wedge\omega$, 
remnants of the singled-fiber bundles $VM$ and $UM$, via the two 
coupled terms $\omega\wedge\varphi$ and $\varphi\wedge\omega$. 
If one transports a \underline{leaf} horizontally along 
a close path on the base-space (in non-flat directions), one measures 
a curvature which is different from that obtained by taking the  
sum of single fiber treks, as one usually does by the process of trivial 
splicing. In fact, the foliar complex can be considered as a 
{\em unifying infrastructure\/} within which two gauges are composed 
into a single structure whose curvature  
$R=d\of{\varphi+\omega}+\of{\varphi+\omega}\wedge\of{\varphi+\omega}$
is made of a {single} connection $\of{\varphi+\omega}$ with values 
in a space product of two Lie algebras, and which satisfies a two-group
implementation of a {single} transformation law,  
$G_{V}:\,\of{\varphi+\omega}\mapsto v\of{\varphi+\omega+d}v^{-1}$,
and $G_{U}:\,\of{\varphi+\omega}\mapsto u\of{\varphi+\omega+d}u^{-1}$.
In this context, the curvature coefficients are given by the more
familiar form, 
$R^{c\C}=d\of{\varphi+\omega}^{c\C}+\frac{1}{2}C_{VU}^{a\A b\B c\C}
\of{\varphi+\omega}_{a\A}\wedge\of{\varphi+\omega}_{b\B}\,$ with the
single-group structure constants now being replaced by the foliar ones.\Par 
We conclude this section by establishing the exact relations between absolute 
differentials and covariant exterior derivatives. To this end we note that if 
$\offf{\bs{e}_{\alpha}}$ stands for a set of frame fields of $VM$ and 
$g^{\alpha\beta}:=g_{\alpha\beta}^{-1}$, where
$g_{\alpha\beta}:=\bs{e}_{\alpha}\cdot\bs{e}_{\beta}$ is a local
metric on the fiber, we have $\bs{e}^{\alpha}:=
g^{\alpha\beta}\bs{e}_{\beta}$, from which
$\bs{e}^{\alpha}\cdot\bs{e}_{\gamma} = g^{\alpha\beta}g_{\beta\gamma}
={\delta^{\alpha}}_{\gamma}$. In contrast, however, the coframe field
monomials of $V^{\star}M$ are generated by the set of
functionals $\offf{\widetilde{\bs{e}}^{\alpha}}$ satisfying
$\widetilde{\bs{e}}^{\alpha}\of{\bs{e}_{\beta}}={\delta^{\alpha}}_{\beta}$.
Therefore, for any $\bs{e}\in VM$, we have $\of{d\bs{e}^{a}}\cdot\bs{e}_{b}=
-\bs{e}^{a}\cdot \of{d\bs{e}_{b}}=
+\bs{e}^{a}\cdot\varphi\of{X}\bs{e}_{b}\,\Rightarrow\, d\bs{e}^{a}=
+\varphi\of{X}\bs{e}^{a}$, and similarly, for any $\bar{\bs{e}}\in UM$, 
$d\bar{\bs{e}}^{\A}=+\omega\of{X}\bar{\bs{e}}^{\A}$, but the coframe
functionals of $V^{\star}M$ and $U^{\star}M$ are all annihilated by $d$ 
because $d\widetilde{\bs{e}}=
\of{\partial\widetilde{\bs{e}}/\partial\bs{e}}d\bs{e}=0$.
Keeping an open eye on the order in which terms are  
arranged in an expression, it immediately follows that 
$d\of{\bs{e}^{a}\otimes\bar{\bs{e}}^{\A}\psi_{a\A}}\equiv
\bs{e}^{a}\otimes\bar{\bs{e}}^{\A}\of{D\psi}_{a\A}$
for vector-valued forms, whereas
$d\of{\bs{e}^{a}\otimes\bar{\bs{e}}^{\A}\psi_{a\A}^{b\B}\otimes\bs{e}_{b}
\otimes \bar{\bs{e}}_{\B}}\equiv \bs{e}^{a}\otimes\bar{\bs{e}}^{\A}
\of{D^{\star}\psi} _{a\A}^{b\B}\otimes\bs{e}_{b}\otimes\bar{\bs{e}}_{\B}$
for tensor-valued ones.
Being a little careless with standards of rigor (by identifying 
differentials taken in different directions) we may take an advantage 
of the ``failure'' of $d\circ d$ to annihilate vector-fields and
further construct the curvature and other objects of interest by 
successive applications of $d$.
\section{\sf{Global Rescaling and Local Translations
in the Spaces of Connections}}\label{s5}
We next observe that the foliar complex construction is defined up to 
multiplicative constants added to the foliar pair of connections; the 
substitutions
\be{19}\ba{llcc}
\varphi\rightarrow a^{-1}\varphi, & \omega\rightarrow b^{-1}\omega & &
a,b\in K
\ea\ee
are compatible with the gauge transformation laws of (\ref{2}) and 
therefore maintain covariance at any level. One may, for example, 
normalize the connections by choosing $a=\int{\cal D}
\off{\varphi}$ and $b=\int{\cal D}\off{\omega}$ where the 
functional integrations are taken in connection
space (modulo gauge transformation) and the ${\cal D}$'s 
are the appropriate measures. 
Nonetheless, each single term in the expressions for the covariant 
exterior derivatives and the curvature is now weighted differently.
Therefore, a global rescaling in each space of connections tunes 
the curvature, thereby reproducing three potential ``self-interaction'' 
couplings (namely, gluon-gluon): $a^{-2}$, $b^{-2}$ and $a^{-1}b^{-1}$,
where of a particular interest is the third one which weights
{\em cross-gauge\/} 
interactions.\Par
Consider next a single-fiber gauge structure, and let the connection 
$\omega$ translate according to $\omega\rightarrow\omega+\Omega$, where
big $\Omega$ is an $x$-dependent ``co-frame'' one-form taking its values 
in the same Lie algebra as $\omega$. By construction, $\Omega$ cannot be
gauged out of the bundle (like a pure gauge) and
therefore $\omega$ and $\omega+\Omega$ can never be connected by a
gauge transformation. Being of a completely horizontal nature, the
translation by $\Omega$ generates bijections 
between equivalence classes in the coset 
space of connections modulo gauge-transformations. Now, while
$\omega$ obviously transforms properly still after being shifted, the 
curvature no longer stays invariant. One is therefore led to introduce
the notion of a 
``vertical'' Grassmann algebra, defined over the group manifold and
graded by an appropriate co-boundary operator $\delta$ with respect to 
which $\Omega$ is considered as one-form as well. Over the extended 
arena of forms, where we allow everything to depend on all possible degrees 
of freedom, $\delta$ anticommutes with $d$. Exploiting the operatorial
definition for the curvature by successively applying the covariant exterior 
derivative to an arbitrary test-form of the extended Grassmann space, we find 
after some calculation: 
\beq{20}
D^{\star}\circ D^{\star}\psi &\!\!=\!\!&
D^{\star}\of{d\psi+\delta\psi
+\of{\omega+\Omega}\wedge\psi+\psi\wedge\of{\omega+\Omega}}\nonumber\\
&\!\!=\!\!& \off{R,\psi}+\off{D^{\star}\Omega,\psi}+\off{\Omega\wedge
\Omega ,\psi}+\off{\delta\of{\omega+\Omega},\psi}.
\eeq 
The vanishing of the extra three commutators in (\ref{20}) uniquely implies 
$\delta\omega=-D^{\star}\Omega$ and $\delta\Omega=-\Omega\wedge\Omega$ 
(uniqueness: a consequence of compatibility with the gradings)
whereas, in the absence of any other conditional terms, the variation 
of $\psi$ is left totally undetermined. One can easily check that a
squared variation $\delta\delta$ vanishes on both $\omega$ and $\Omega$.
Two subsidiary comments are in order: 
First, the indifference of the curvature to
horizontal translations in the space of connections, induced by the
above structure, just manifests a known ambiguity 
due to Wu and Yang, where two 
gauge-disconnected non-Abelian connections give rise to the same curvature.  
Second, the coefficients $\Omega^{a}$ (and later on $\Omega^{a\A}$ and
$\Phi^{a\A}$ as well), being differential forms, vanish upon exterior 
squaring and can therefore be identified with the {\em ghosts\/} of
the gauge. The following digression concludes our single-structure prelude;
\Par{\em Digression\/}:
According to the conventional geometric approach to ghost sectors 
and BRST co-homology \cite{3}, one enlarges the classical
Grassmann bundle to include the vertical space spanned by the group angles: 
Pick a local coordinate frame $x\in M,\;\phi\of{x}\in G$, and put 
$\ac{\omega}=\omega^{a}_{\mu}\of{x,\phi\of{x}}L_{a} dx^{\mu}+C^{a}_{b}
\of{x,\phi\of{x}}
L_{a}\delta\phi^{\,b} \of{x}$, where the $L_{a}$'s $\in\mbox{Lie}\,G$
and the $C_{b}$'s stand for the vertical components of the connection.
Note: we now deal with an extended base in which for any $g\in G$, 
$\ac{\omega}\mapsto g\of{\ac{\omega}+d+\delta}g^{-1}$. 
Abusing slightly the notations, the corresponding two co-boundary operators
are realized by $d:=dx^{\mu}\partial/\partial x^{\mu}$ and $\delta:=
d\phi^{a}\partial/\partial \phi^{a}$, where the realizations in particular
imply $d\circ d=\delta\circ\delta=d\circ\delta+\delta\circ d=0$, and
moreover, $\delta\phi^{a}=d\phi^{a}$. As a consequence of the above
extension, an expansion of the corresponding curvature results in
four terms, 
$R_{\ms{extended}}=R_{\ms{hh}}+R_{\ms{hv}}+R_{\ms{vh}}+R_{\ms{vv}}$, where 
$R_{\ms{hh}}$ is a purely horizontal form, $R_{\ms{vv}}$ is a purely
vertical one, and  $R_{\ms{hv}}$, $R_{\ms{vh}}$ are forms of
a mixed basis. Then, the variations of $\omega$ and $\Omega$ 
with respect to $\delta$ follow by requiring flatness in vertical 
directions, namely, by imposing $R_{\ms{hv}}=R_{\ms{vh}}=R_{\ms{vv}}=0$. 
In this context, the gauge sector is said to possess an internal 
BRST structure. 
Now, $G$ is a smooth manifold $\Rightarrow$ $d\phi^{a}=
\of{\partial\phi^{a}/\partial x^{\mu}}dx^{\mu}$ $\Rightarrow$ $\ac{\omega}$
employs base-space components only. This, in turn, exactly relates our 
$\Omega^{a}$'s to the ghosts of the above description, namely, 
$\Omega^{a}\equiv C^{a}_{b}(\partial\phi^{b}/\partial
x^{\mu})dx^{\mu}$, where $\ac{\omega}$ is
just our original $\omega$ that has been shifted by $\Omega$. The idea is
easily extended to forms of arbitrary degree where
the ghost index counts the number of $\of{\partial\phi/\partial x}$'s
occurring in a base space implementation, {\em not\/} the 
vertical gradation. It is therefore why a connection
$\omega$, as oppose to a shift $\Omega$, carries ghost index zero.\Par
Let us now get back to foliar complexes. In this case,  
each of the two connection forms $\omega$ and $\varphi$ is shifted in
its own coset space by a corresponding ``co-frame'' one-form (say, big
$\Omega$ and big $\Phi$ respectively), taking values in $\mbox{Lie}\,G_{V}
\otimes\mbox{Lie}\,G_{U}$. Under these circumstances, both shifted terms
transform precisely according to eqs. (\ref{2}). In order to fix the
curvature, we let the connections and the shifted terms depend on vertical
variables as well. This time, however, there are two sets of them, 
corresponding to the two group manifolds spanned by $G_{V}$ and $G_{U}$. 
Each of the associated Grassmann spaces is graded by its own 
co-boundary operator (denoted respectively by $\delta$ and $\bar{\delta}$)
with respect to which $\Omega$ and $\Phi$ are considered as one forms. 
One now encounters two ghost indexes: one is generated by $\delta$, the 
other is generated by $\bar\delta$; a single quantum of the former is 
carried by $\Omega$, a single quantum of the latter, by $\Phi$. 
And of course, $d\delta+\delta d = d\bar\delta+\bar\delta d = \delta\bar
\delta+\bar\delta\delta = 0$. Making use of the operatorial definition 
for the curvature with respect to a base space of $n\times n_{V}\times n_{U}$
dimensions, we find:
\be{22}\ba{lcl}
D^{\star}\circ D^{\star}\psi &\!\!=\!\!&
\off{R,\psi}+\off{D^{\star}\of{\Omega+\Phi},\psi}+\off{\of{\Omega+\Phi}
\wedge\of{\Omega+\Phi},\psi}\\
&& +\:\off{\delta\of{\omega+\Omega+\varphi+\Phi},
\psi}+\off{\bar{\delta}\of{\omega+\Omega+\varphi+\Phi},\psi}. 
\ea\ee
Once more, we set the extra four terms to vanish by equating 
the terms of equal ghost index. The resulted variation laws 
read:
\be{24}\ba{lcl}
\delta\of{\omega+\varphi}=-D^{\star}\Omega&&
\bar\delta\of{\omega+\varphi}=-D^{\star}\Phi\\
\delta\Omega=-\Omega\wedge\Omega,&&\bar\delta\Omega=-\Phi\wedge\Omega,\\
\delta\Phi=-\Omega\wedge\Phi,&&\bar\delta\Phi=-\Phi\wedge\Phi.\\
\ea\ee
We see that we cannot derive separate variation laws for $\varphi$ and
$\omega$. This, however, is totally compatible with our previous
claim where we argued that $\varphi+\omega$ can be regarded as a 
single connection of a single gauge structure with two gauge groups; 
the BRST variations see only one gauge connection.\Par
It is manifestly evident that (\ref{24}) is completely
invariant with respect to a duality transformation, 
$\delta\leftrightarrow\bar\delta$ and $\Omega\leftrightarrow\Phi$. 
In addition, one easily verifies that the two squared variations,
$\delta\delta$ and $\bar\delta\bar\delta$,
vanish on $\Omega$, $\Phi$ and $\omega+\varphi$. Now,
the information about how ghosts vary with respect to the 
co-boundary operators of the counterpart Grassmann space can 
be compactly described by 
$\delta\Phi+\bar\delta\Omega+\off{\Omega,\Phi}=0$, see (\ref{24}).
Guided by such a duality-invariant relation, let us introduce
$B:=\delta\Phi=-\Omega\wedge\Phi$, by construction
$\delta$-exact, an entity whose coefficients in $\mbox{Lie}\,G_{V}
\otimes\mbox{Lie}\,G_{U}$ are commuting differential forms.
Then, due to the nilpotency of $\delta$ and due to (\ref{24}),
\be{24a}\ba{lclcl}
\delta B=0,&&\bar\delta B=-\off{\Phi,B}&\Rightarrow&\bar\delta\bar\delta B=0.
\ea\ee
The structural and variation properties of $B$ suggest to identify
it with the so-called $B$-field in the context of the BRST mechanism. 
However, in contrast with previous treatments \cite{3,4}, 
here it appears not as additional degrees of 
freedom, but rather as a simple composition of already existing
ones, namely, a ghost-ghost exterior product. 
Had we started, however, with an alternative assignment, 
$\bar{B}:=\bar\delta\Omega=-\Phi\wedge\Omega$,
we would have obtain the same results, only within the dual description.
It therefore makes sense to identify the dual $\bar{B}$-field 
with the anti-field of $B$. Consider for a moment the case of a unitary
structure: Being just shifts between hermitian fields, $\Phi$ and $\Omega$
are themselves hermitian, and we have $B^{\dagger}=-\bar{B}$. 
Following this line of reasoning, $\Phi$ and $\Omega$ appear as 
anti-fields of one another. We can therefore pick one of them, 
no matter which, to play the role of an anti-ghost. 
These interpretations, together with the derived variation 
laws above, are seen to cover
the entire BRST structure of a foliar complex,
all by pure geometrical means. Apparently, it 
coincides with the ghost-anti-ghost super algebra
one usually associates with local gauge theories.
 
\section{\sf{Splices Which Admit Contractable Pieces}} 
\label{s6}
Let us now soften the constraints for closed anti-commutability
in an appropriate representation and include identity elements as well.
Namely, the generators of $G_{V}$, while carrying representations in
$V$, satisfy $\offf{L^{a},L^{b}}=\frac{2}{N_{V}}d_{V}^{ab}+g_{V}^{abc}L_{c}$, 
where $N_{V}=\mbox{dim}\,V$, $d_{V}^{ab}=\mbox{Tr}\of{L^{a}L^{b}}$ 
and similar relations hold for the $\LL$'s of $G_{U}$ with the corresponding
$d_{U}$'s, and $N_{U}=\mbox{dim}\,U$. (Note: this is exactly the case
of two $\mbox{SU}\of{n}$ structures in their fundamental representation).
Now, in order to comply with the 
appearance of the identity, one first prolongs the vector spaces spanned by 
the generating algebras by adding $L_{0}=\LL_{0}:=I$, and then extends the 
closure formulae (\ref{3}) to include (we use $\alpha,\beta,
\ldots=0,1,\ldots,n_{V}$ and $\Alpha,\Beta,\ldots=0,1,\ldots,n_{U}$): 
\be{30}\ba{lll}
g_{V}^{\alpha\beta 0}=\frac{2}{N_{V}}d_{V}^{\alpha\beta}&&
g_{V}^{\alpha 0\gamma}=2\delta^{\alpha\gamma}\\
g_{U}^{\Alpha\Beta 0}=\frac{2}{N_{U}}d_{U}^{\Alpha\Beta}&&
g_{U}^{\Alpha 0\GGamma}=2\delta^{\Alpha\GGamma},\ea
\ee
where for the $f$'s we put $f_{V}^{\alpha\beta\gamma}=0$ if 
\underline{any} of the indices $\alpha,\beta$ or $\gamma$ is zero, and  
the same for $f_{U}^{\Alpha\Beta\GGamma}$ with $\Alpha,
\Beta$ and $\GGamma$. Note that the $d^{\alpha\beta}$'s and the
$d^{\Alpha\Beta}$'s are now interpreted as Cartan metrics on the unitals 
(unital $=$ a Lie algebra $+$ the unity).
Let us extract the consequences of this 
generalization; the original connection one forms, in principle, are
still valued in $\mbox{Lie}\,G_{V}\otimes\mbox{Lie}\,G_{U}$. 
Introduce, however, an extra pair of scalar one forms 
$\of{\varphi_{00}L^{0}\otimes\LL^{0},\omega_{00}L^{0}\otimes\LL^{0}}$,
an extra pair of single-fiber connections
$\of{\varphi_{a0}L^{a}\otimes\LL^{0},\omega_{0\A}L^{0}\otimes\LL^{\A}}$
and a pair of single-fiber tensor one-forms
$\of{\varphi_{0\A}L^{0}\otimes\LL^{\A},\omega_{a0}L^{a}\otimes\LL^{0}}$
and put $\widehat\omega=\omega_{\alpha\Alpha}L^{\alpha}\otimes\LL^{\Alpha}$, 
and $\widehat\varphi=\varphi_{\alpha\Alpha}L^{\alpha}\otimes\LL^{\Alpha}$. 
\Par\noindent{\em Comment\/}: Namely,
$\varphi_{a0}L^{a}\otimes\LL^{0}$ is a connection form on $VM$, 
but $\varphi_{0\A}L^{0}\otimes\LL^{\A}$ is a rank-2 $G_{U}$-tensor 
form on $UM$; $\omega_{0\A}L^{0}\otimes\LL^{\A}$ is a connection 
form on $UM$, but $\omega_{a0}L^{a}\otimes\LL^{0}$ is a rank-2 
$G_{V}$-tensor form on $VM$.\Par
Now, since $\widehat\varphi$ and $\widehat\omega$ are actually valued in 
$\of{\mbox{Lie}\,G_{V}+I_{V}}\otimes\of{\mbox{Lie}\,G_{U}+I_{U}}$,
any symmetric product of two `hats', like $\widehat\omega\wedge\widehat\omega$
or $\off{\widehat\varphi,\widehat\omega}$, necessarily involves a 
structural classification according to $\mbox{Lie}\,G_{V}\otimes\mbox{Lie}\,
G_{U}+\mbox{Lie}\,G_{V}\otimes I_{U}+I_{V}\otimes\mbox{Lie}\,G_{U}+I_{V}
\otimes I_{U}$. We notice that the trivial part of the splice is present 
via $\mbox{Lie}\of{G_{V}\times G_{U}}$. Tracing the recipes given by the 
proposition of section \ref{s3}, we find the following expression for  
the total ``curvature'' two-form $\widehat R$:
\beq{31}
&&\widehat{R}\;:=\;2\times d\widehat\omega+\widehat\omega\wedge\widehat\omega+
\of{\widehat\omega\wedge\widehat\varphi+\widehat\varphi\wedge\widehat\omega}
+\widehat\varphi\wedge\widehat\varphi+2\times d\widehat\varphi \\
&&\;=\sum_{\varpi,\varpi'=\widehat\varphi,\widehat\omega}
\of{2d\varpi_{\gamma\GGamma}+
\frac{1}{2}\widehat{C}_{VU\;\;\,\gamma\GGamma\,}
^{\alpha\Alpha\beta\Beta}\varpi_{\alpha\Alpha}\wedge
\varpi'_{\beta\Beta}}L^{\gamma}\otimes\LL^{\GGamma}\nonumber
\eeq
with $\widehat{C}_{VU}^{\alpha\Alpha\beta\Beta\gamma\GGamma}
=\frac{1}{2}\of{f_{V}^{\alpha\beta\gamma}g_{U}^{\Alpha\Beta\GGamma}
+f_{U}^{\Alpha\Beta\GGamma}g_{V}^{\alpha\beta\gamma}}+
f_{V}^{\alpha\beta\gamma}f_{U}^{\Alpha\Beta\GGamma}$. (The exact 
differentials above appear twice just in order to account for
linearity; we shall soon discuss this over.) The structural
classification of (\ref{31}) is now dictated by the $\gamma\GGamma$-pairs:
the $\widehat{C}_{VU}$-attached $00$-pair obviously vanishes 
(because each term in $\widehat{C}_{VU}$ has an $f$ representative), 
the $c0$-pairs and the $0\C$-pairs, 
each correspond to a single-fiber structure, whereas the $c\C$-pairs 
correspond to a foliar-like structure.\Par
One should, however, be very cautious with gauge interpretations: 
Almost non of these decomposed terms vary
linearly with respect to gauge transformations. What then are the
transformation properties of $\widehat{R}$\,? 
Before answering this question, let us adopt the following
short-hand notation: $\widehat\varphi=
\varphi_{00}L^{0}\otimes\LL^{0}+\varphi_{a0}L^{a}\otimes\LL^{0}+\varphi_{0\A}
L^{0}\otimes\LL^{\A}+\varphi_{a\A}L^{a}\otimes\LL^{\A}
:=\varphi^{00}+\varphi^{10}+\varphi^{01}+\varphi^{11}$, 
and the same for $\widehat\omega$.
Then, for example, $\varphi^{10}$ is a $G_{U}$-scalar connection
one-form on $VM$, $d\varphi^{10}+\varphi^{10}\wedge\varphi^{10}$ 
is a $VM$-curvature element
$\in R^{1010}=d\varphi^{10}+\varphi^{10}\wedge\varphi^{10}+\varphi^{10}
\wedge\omega^{10}+\omega^{10}\wedge\varphi^{10}+\omega^{10}\wedge\omega^{10}
+d\omega^{10}$, whereas $\varphi^{01}\wedge\varphi^{01}$ is a pure
$G_{U}$-tensor $\in R^{0101}=d\varphi^{01}+\varphi^{01}\wedge\varphi^{01}
+\varphi^{01}\wedge\omega^{01}+\omega^{01}\wedge\varphi^{01}+\omega^{01}
\wedge\omega^{01}+d\omega^{01}$ etc. In addition we put 
$R^{1111}=d\varphi^{11}+\varphi^{11}
\wedge\varphi^{11}+\varphi^{11}\wedge\omega^{11}+\omega^{11}\wedge\varphi^{11}
+\omega^{11}\wedge\omega^{11}+d\omega^{11}$, but note that 
this object is of a mixed algebraic structure because the wedge
operation now activates identity elements as well; it however 
transforms linearly with respect to both gauge groups.
Now, as the quantity $\widehat{R}$ is totally symmetric
with respect to its two arguments, $\varphi$ and $\omega$, 
all of its wedge products that involve
scalar forms (there are 36 of them) vanish identically, and moreover,
any of the four symmetric pairs $\of{1001}$ and $\of{0110}$ 
vanishes as well because each wedge product in such a pair
consists of two commuting terms, each of which living in a different
representation space. Therefore, and with the above 
conventions at our disposal, $\widehat{R}$ decomposes into
\be{70}\ba{lcl}
\widehat{R}&=&R^{1010}+d\varphi^{10}+d\omega^{10}
+R^{0101}+d\varphi^{01}+d\omega^{01}+
R^{1111}+d\varphi^{11}+d\omega^{11}\\
&&+\;2d\omega^{00}+2d\varphi^{00}+\;\off
{\varphi^{10}+\varphi^{01}+\omega^{10}+\omega^{01},\varphi^{11}+\omega^{11}}
\ea\ee
(the last term is a graded commutator). It is straightforward, and not too
cumbersome, to verify the following gauge transformation properties of
$\widehat{R}$:
\be{71a} 
G_{V}:\left\{\ba{l}\left.\ba{l}
R^{1010}+R^{1111}+d\varphi^{10}+
d\omega^{10}+d\varphi^{11}+d\omega^{11}\\
+\;\off{\varphi^{10}+\varphi^{01}+\omega^{10}+\omega^{01},
\varphi^{11}+ \omega^{11}},\ea \right\}
\;\;\Longrightarrow\;\;\mbox{$G_{V}$-tensor} \\
\;\;R^{0101}+ 2d\varphi^{00}+d\varphi^{01}+2d\omega^{00}+d\omega^{01}
\;\;\Longrightarrow\;\;\mbox{$G_{V}$-scalar}\ea\right.
\ee
\be{71b}
G_{U}:\left\{\ba{l}\left.\ba{l}
R^{0101}+R^{1111}+d\varphi^{01}+d\omega^{01}+d\varphi^{11}+d\omega^{11}\\
+\;\off{\varphi^{10}+\varphi^{01}+\omega^{10}+\omega^{01},
\varphi^{11}+\omega^{11}},\ea \right\}
\;\;\Longrightarrow\;\;\mbox{$G_{U}$-tensor}\\
\;\;R^{1010}+ 2d\varphi^{00}+d\varphi^{10}+2d\omega^{00}+d\omega^{10}
\;\;\Longrightarrow\;\;\mbox{$G_{U}$-scalar}.
\ea\right.\ee\Par\noindent
{\em Comment\/}: In fact,
$\of{d\varphi^{10}+d\varphi^{11}+d\omega^{10}+d\omega^{11}}
+\;\off{\varphi^{10}+\varphi^{01}+\omega^{10}+\omega^{01},
\varphi^{11}+ \omega^{11}}$, $R^{1010}$ and $R^{1111}$, each of which
behaves as an independent $G_{V}$-covariant quantity. In the same manner,
$\of{d\omega^{01}+d\varphi^{11}+d\omega^{01}+d\omega^{11}}
+\off{\varphi^{10}+\varphi^{01}+\omega^{10}+\omega^{01},
\varphi^{11}+ \omega^{11}}$, $R^{0101}$ and $R^{1111}$,  
all behave as independent $G_{U}$-covariant quantities.\Par
In conclusion, $\widehat{R}$ possesses simple but non-trivial
transformation properties. In particular, it admits terms that
behave as scalars under various group actions, each one at a time. 
One should not, however, be too much surprised: One now 
deals with structures that are induced by central extensions of 
Lie algebras rather than the Lie algebras themselves. In the former
case, as opposed to the latter one, one starts with a ``curvature'' 
which inherently contains scalar pieces (with respect to one gauge 
group or the other, or both) and therefore one ends-up with scalar terms.
This fact, in general, leads to the lack of covariance for the case were
sectors are classified according to their algebraic structure, where there
is only one exception: There are two autonomic decoupled sectors which
are linear, and whose definite algebraic structure is completely
preserved by the transformations, namely, the two single fiber
bundles whose composition is given just by trivial means; 
\beq{80}
R^{1010}&\!\!=\!\!\!&
\sum_{\varpi,\varpi'=\omega,\varphi}\of{d\varpi^{c0}+
\frac{1}{2}f_{V}^{abc}\varpi_{a0}\wedge\varpi'_{b0}}
L_{c}\otimes I_{U}\;\,\in VM \\
R^{0101}&\!\!=\!\!\!&
\sum_{\varpi,\varpi'=\omega,\varphi}\of{d\varpi^{0\C}+
\frac{1}{2}f_{U}^{\A\B\C}
\varpi_{0\A}\wedge\varpi'_{0\B}}I_{V}\otimes\LL_{\C}\;\,\in UM.
\eeq
These two pieces closed on themselves and therefore can be treated
independently of everything else. In this case, single-structure
bundle operators will certainly do. Otherwise, if everything is taken
into account, one has to utilize modified bundle
operators which generalize the foliar complex covariant exterior
derivatives. In particular, the appropriate expression for a covariant
exterior derivative of vector-valued leaf forms reads: 
$\widehat{D}\psi := 2d\psi+\of{\widehat\varphi+\widehat\omega}
\wedge\psi$ while that of tensor valued leaf forms,
$\widehat{D}^{\star}\psi := 2d\psi +
\of{\widehat\varphi+\widehat\omega}\wedge\psi + \of{-1}^{p+1}
\psi\wedge\of{\widehat\varphi+\widehat\omega}$ where
$p=\mbox{deg}\,\psi$. One then treks 
familiar pathways, first by reconstructing the ``curvature'' via
$\off{\widehat{D},\widehat{D}}\psi=2\widehat{R}\wedge\psi$, 
or alternatively via
$\off{\widehat{D}^{\star},\widehat{D}^{\star}}\psi:=2\off{\widehat{R},\psi}$,
and later constructing identities by successive applications of
$\widehat{D}$ or $\widehat{D}^{\star}$. This explains why
the scalar one-form terms that are present in $\widehat{D}$ and
$\widehat{D}^{\star}$ are indeed integrated pieces of the model and
cannot be dropped out; they do not count for linearity but for
the consistency and completeness sake.
\section{Epilogue}
We have seen that nothing indeed prevents us from generalizing the
concept of a foliar complex, just by going to central extensions. Other
extensions, however, are believed to be completely acceptable as long as  
anti-commutability can be softly expressed in a closed form;
that is, terms added to the algebra should not necessarily be 
proportional to the identity provided they can always be incorporated in 
the generalized form of eq. (\ref{3}) where the indices run over
the prolonged vector space. As a result, one trades with the overall 
linearity and the pure foliar structure is lost. Instead      
one gets two autonomic, mutually transparent, single fiber structures
embedded in wider frames in which some of the original foliar
properties are still inherent.\par
It would seem that a generalization of the whole model
to the case of composing {\em any\/} number of gauge structures is
straightforward: For an $r$-fold composition with $r$ gauge
groups, introduce $r$ connection one-forms with values in the product
space of the corresponding $r$ Lie algebras, each of which transforms
as a connection with respect to its inducing gauge group and as a tensor with
respect to all the other ones. The pure foliar complex is then read-off 
from a curvature of the type $R=d\of{\varphi_{1}+\cdots+\varphi_{r}}+
\of{\varphi_{1}+\cdots+\varphi_{r}}\wedge\of{\varphi_{1}+\cdots+\varphi_{r}}$,
and the two types of 
covariant exterior derivatives of $r$-fold foliar forms
are made of an $r$-connection one-form $\of{\varphi_{1}+\cdots+\varphi_{r}}$.
Such a large structure will admit an $r$-fold BRST symmetry with
associated $r$ ghosts. Once more, extensions at the level of the algebra
break the pure foliar structure at the gain of two decoupled ones. 
The generalization idea, however, is still speculative and requires a 
closer examination.\par
Finally, we mark on the symmetric role played by the two gauge connections: 
In a quantum theory, since both multiplets involve the same representation
under both groups, the symmetry becomes a source 
for degeneracies. In particular, states of the gauge sector are
expected to mix with one another, and the phenomenology is no longer
automatically compatible with the gauge structure, especially if 
some mechanism is added in order to break the original gauge symmetry.
\Par\Par\noindent
{\large\sf{Acknowledgments}}:
I thank Yuval Ne'eman for important comments, L. P.
Horwitz for reading an early (and partial) version of these notes.
I am also indebted to S. Sternberg for critically commenting on 
some early preliminary definitions. The work was partially 
supported by the Ann \& Maurice Jacob Cohen Doctoral 
Fellowship in Nuclear and Particle Physics.

\end{document}